\documentclass[9pt,twocolumn]{extarticle}
\usepackage[margin=0.75in]{geometry}
\usepackage{amssymb}
\usepackage{amsfonts}
\usepackage{amsmath}
\usepackage{authblk}
\usepackage{blindtext}
\usepackage{graphicx}
\usepackage{color}
\usepackage{widetext}
\usepackage[numbers,sort&compress]{natbib}
\usepackage[scientific-notation=true]{siunitx}

\makeatletter
\newcommand{\printfnsymbol}[1]{%
  \textsuperscript{\@fnsymbol{#1}}%
}
\makeatother

\makeatletter
\newbox\abstract@box
\renewenvironment{abstract}
  {\global\setbox\abstract@box=\vbox\bgroup
     \hsize=\textwidth\linewidth=\textwidth
    \vspace{-0.5cm}
    \begin{center}%
    {\bfseries \abstractname\vspace{-.5em}\vspace{\z@}}%
    \end{center}%
    \quotation}
  {\endquotation\egroup}
\expandafter\def\expandafter\@maketitle\expandafter{\@maketitle
  \ifvoid\abstract@box\else\unvbox\abstract@box\if@twocolumn\vskip1.5em\fi\fi}
\makeatother

\makeatletter
\newcommand\footnoteref[1]{\protected@xdef\@thefnmark{\ref{#1}}\@footnotemark}
\makeatother
\newcommand{\pa}{\partial}

\newcommand{\pd}[2]{\frac{\partial #1}{\partial #2}}
\newcommand{\pdd}[2]{\frac{\partial^2 #1}{{\partial #2}^2}}
\newcommand{\pddd}[2]{\frac{\partial^3 #1}{{\partial #2}^3}}

\usepackage{etoolbox} 
\pretocmd{\eqref}{Eq.~}{}{}

\usepackage[usenames]{xcolor}
\usepackage{hyperref}
\hypersetup{
	colorlinks,
	linkcolor={blue!50!black},
	citecolor={blue!50!black},
	urlcolor={blue!80!black}
}
\renewcommand{\O}{\mathcal{O}}

\title{Learning data driven discretizations for partial differential equations}

\author[1]{Yohai Bar-Sinai\thanks{YBS and SH contributed equally to this work}\thanks{ybarsinai@gmail.com}}
\author[2]{Stephan Hoyer\printfnsymbol{1}\thanks{shoyer@google.com}}
\author[2]{Jason Hickey}
\author[1,2]{Michael P.~Brenner}

\affil[1]{School of Engineering and Applied Sciences, Harvard University, Cambridge, MA}
\affil[2]{Google Research, 1600 Amphitheatre Pkwy, Mountain View CA 94043}

\date{}
\begin{document}

\begin{abstract}
The numerical solution of partial differential equations (PDEs) is challenging because of the need
to resolve spatiotemporal features over wide length and timescales.
Often, it is computationally intractable to resolve the finest features in the solution.
The only recourse is to use approximate coarse-grained representations,  which aim to accurately represent long-wavelength dynamics while properly accounting for unresolved small scale physics.
Deriving such coarse grained equations is notoriously difficult, and often \emph{ad hoc}.
Here we introduce \emph{data driven discretization}, a method for learning optimized approximations to PDEs based on actual solutions to the known underlying equations.
Our approach uses neural networks to estimate spatial derivatives, which are optimized end-to-end to best satisfy the equations on a low resolution grid.
The resulting numerical methods are remarkably accurate, allowing us to integrate in time a collection of nonlinear equations in one spatial dimension at resolutions 4-8x coarser than is possible with standard finite difference methods.
\end{abstract}

\maketitle

Solutions of nonlinear partial differential equations can have enormous complexity, with nontrivial structure over a large range of length and timescales. Developing effective theories that integrate out short length scales and fast time scales is a long standing goal.
As examples, geometric optics is an effective theory of Maxwell equations at scales much longer than the  wavelength of light~\cite{Jackson1999}; Density Functional Theory models the full many-body quantum wavefunction with a lower dimensional object -- the electron density field~\cite{DFT_Sholl}; and the effective viscosity of a turbulent fluid parametrizes how  small scale features affect large scale behavior~\cite{ChenTurbulence}.
These models derive their coarse-grained dynamics by more or less systematic integration of the underlying governing equations (by using, respectively, WKB theory, Local Density Approximation and a closure relation for the Reynold stress).
The gains from coarse graining are, of course, enormous.
Conceptually, it allows a deep understanding of emergent phenomena that would otherwise be masked by irrelevant details.  Practically, it allows computation of vastly larger systems.

Averaging out unresolved degrees of freedom  invariably replaces them by  effective parameters that mimic typical behavior.
In other words, we identify the salient features of the dynamics at short-and-fast scales and replace these with terms that have a similar average effect on the long-and-slow scales.
Deriving reliable effective equations is often challenging~\cite{Van-Dyke1975}.
Here we approach this challenge from the perspective of statistical inference.
The coarse-grained representation of the function contains only partial information about it, since short scales are not modeled.
Deriving coarse-grained dynamics requires first inferring the small scale structure using the partial information (reconstruction) and then  incorporating its effect on the coarse-grained field.
We propose to perform reconstruction using machine-learning algorithms, which have become extraordinarily efficient at identifying and reconstructing recurrent patterns in data.
Having reconstructed the fine features, modeling their effect can be done using our physical knowledge about the system.
We call our method \emph{data-driven discretization}.
It is qualitatively different from coarse graining techniques that are currently in use:
instead of analyzing equations of motion to derive effective behavior, we directly learn from high resolution solutions to these equations.

\paragraph*{Related work}

Several related approaches for computationally extracting effective dynamics have been previously introduced.
Classic works used neural networks for discretizing dynamical systems~\cite{kevrekidis1998, RicoKevrekidis}.
Similarly, equation-free modeling approximates coarse-scale derivatives by remapping coarse initial conditions to fine scales which are integrated exactly~\cite{Kevrekidis2009}.
The method has similar spirit to our approach, but it does not learn from fine-scale dynamics and use the memorized statistics in subsequent times to reduce the computational load.
Recent works have applied machine learning to PDEs, either focusing on speed~\cite{Kim2018, Tompson2017, Rasp2018} or recovering unknown dynamics~\cite{Brunton2016, Bezenac2018}.
Models focused on speed often replace the slowest component of a physical model with machine learning, e.g., the solution of Poisson's equation in incompressible fluid simulations~\cite{Tompson2017}, sub-grid cloud models in climate simulations~\cite{Rasp2018}, or building reduced order models that approximate dynamics in a lower dimensional space~\cite{Lusch2017, Morton2018, Kim2018}.
These approaches are promising, but learn higher-level components than our proposed method.
An important development is the ability to satisfy some physical constraints exactly by plugging learned models into a fixed equation of motion.
For example, valid fluid dynamics can be guaranteed by learning either velocity fields directly~\cite{Bezenac2018} or a vector potential for velocity in the case of incompressible dynamics~\cite{Kim2018}.
Closely related to this work, neural networks can be used to calculate closure conditions for coarse grained turbulent flow models~\cite{Ling2016, Beck2018}.
However, these models rely on existing coarse-grained schemes specific to turbulent flows and do not discretize the equations directly.
Lastly, \cite{Roberts2001} suggested discretizations whose solutions can be analytically guaranteed to converge to the center manifold of the governing equation, but not in a data-driven manner.

\section{Data driven sub-grid scale modeling}

Consider a generic PDE, describing the evolution of a continuous field $v(x,t)$
\begin{equation}
\frac{\pa v }{\pa t}=F\!\left(t,x, v, \pd{v}{x_i}, \frac{\pa v}{\pa x_i \pa x_j}, \cdots \right)\ .
\label{eq:generic_PDE}
\end{equation}
Most PDEs in the exact sciences can be cast in this form, including equations that describe hydrodynamics, electrodynamics, chemical kinetics and elasticity.
A common algorithm to numerically solve such equations is the method of lines~\cite{Schiesser1991}: given a spatial discretization $x_1,\cdots,x_N$, the field $v(x,t)$ is represented by its values at node points $v_i(t)=v(x_i,t)$ (finite differences), or by its averages over a grid cell, $v_i(t)=\Delta x^{-1}\int_{x_i-\Delta x/2}^{x_i+\Delta x/2}v(x',t)dx'$  (finite volumes), where $\Delta x=x_i-x_{i-1}$ is the spatial resolution~\cite{Leveque}.
The time evolution of $v_i$ can be computed directly from \eqref{eq:generic_PDE} by approximating the spatial derivatives at these points.
There are various methods for this approximation -- polynomial expansion, spectral differentiation, etc.\ -- all yielding formulas resembling
\begin{equation}
\frac{\pa^n v}{\pa x^n}\approx \sum_i \alpha_i^{(n)} v_i\ ,
\label{eq:FD}
\end{equation}
where the $\alpha_i^{(n)}$ are precomputed coefficients.
For example, the one dimensional (1D) finite difference approximation for $\pd{v}{x}$ to first-order accuracy is $\partial_x v(x_i)=\frac{v_{i+1}-v_{i}}{\Delta x}+\O(\Delta x)$.

Standard schemes use one set of pre-computed coefficients for all points in space, while more sophisticated methods alternate between different sets of coefficients according to local rules~\cite{Harten1987, WENO_lecture_notes}.
This discretization transforms~\eqref{eq:generic_PDE} into a set of coupled ordinary differential equations of the form
\begin{equation}
\frac{\pa v_i }{\pa t}=F\left(t,x, v_1, \cdots, v_N\right)\ ,
\label{eq:lines}
\end{equation}
that can be numerically integrated using standard techniques.
The accuracy of the solution to \eqref{eq:lines}  depends on $\Delta x$, converging to a solution of \eqref{eq:FD} as $\Delta x\to 0$. Qualitatively, accuracy requires that $\Delta x$ is smaller than the spatial scale of the smallest feature of the field $v(x,t)$ .

However, the scale of the smallest features is often orders of magnitude smaller than the system size.
High performance computing has been driven by the ever increasing need to accurately resolve smaller scale features in PDEs.
Even with petascale computational resources, the largest direct numerical simulation of a turbulent fluid flow ever performed has Reynolds number of order 1,000, using about $5\times 10^{11}$ grid points~\cite{Lee2015,Clay2017, Iyer2017}.
Simulations at higher Reynolds number require replacing the physical equations with effective equations that model the unresolved physics.
These equations are then discretized and solved numerically, e.g., using the method of lines.
This overall procedure essentially modifies~\eqref{eq:FD}, by changing the $\alpha_i$ to account for the unresolved degrees of freedom, replacing the discrete equations in~\eqref{eq:lines} with a different set of discrete equations.

The main idea of this work is that unresolved physics can instead be learned directly from data.
Instead of deriving an approximate coarse-grained continuum model and discretizing it, we suggest directly learning low-resolution discrete models that encapsulate unresolved physics.
Rigorous mathematical work shows that the dimension of a solution manifold for a nonlinear PDE  is finite~\cite{constantin2012,foias1988}, and that approximate parameterizations can be constructed~\cite{jolly1990, titi1990, marion1989}.
If we knew the solution manifold we could generate \emph{equation specific} approximations for the spatial derivatives in~\eqref{eq:FD}, approximations that have the potential to hold even when the system is under-resolved.
In contrast to standard numerical methods, the coefficients $\alpha_i^{(n)}$ are equation-dependent. Different regions in space (e.g., inside and outside a shock) will use different coefficients.
To discover these formulae, we use machine learning: we first generate a training set of high resolution data, and then learn the discrete approximations to the derivatives in \eqref{eq:FD} from this dataset.
This produces a trade off in computational cost, which can be alleviated by carrying out high resolution simulations on small systems to develop local approximations to the solution manifold, and using them to solve equations in much larger systems at significantly reduced spatial resolution.

\paragraph*{Burgers' equation}

For concreteness, we demonstrate this approach with a specific example in one spatial dimension.
Burgers' equation is a simple nonlinear equation which models fluid dynamics in 1D and features shock formation. In its conservative form, it is written as:
\begin{equation}
\pd{v}{t} + \frac{\partial }{\partial x}J\left(v,\pd{v}{x}\right) = f(x,t)\ ,
\quad
J\equiv \frac{v^2}{2} - \eta \pd{v}{x},
\label{eq:burgers}
\end{equation}
where $\eta > 0$ is the viscosity and $f(x,t)$ is an external forcing term. $J$ is the called a \textit{flux}.
Generically, solutions of \eqref{eq:burgers} spontaneously develops sharp shocks, with specific relationships between the shock height, width and velocity~\cite{Leveque} that define the local structure of the solution manifold.

\begin{figure}
	\centering
	\includegraphics[width=\columnwidth]{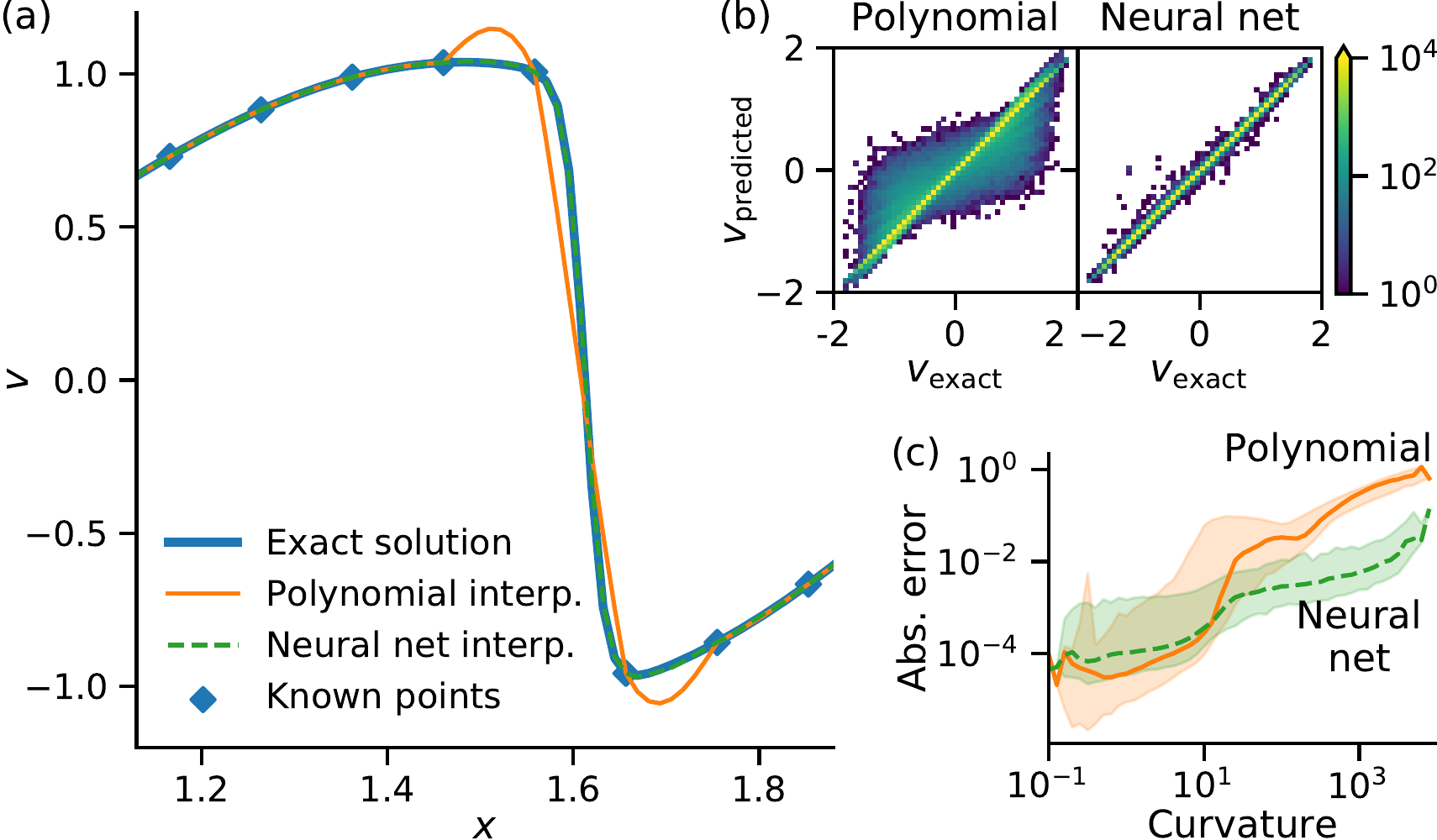}
	\caption{
		\textbf{Polynomial vs.\ neural net based interpolation}
		(a) Interpolation between known points  (blue diamonds) on a segment of a typical solution of Burgers' equation.
		Polynomial interpolation exhibits spurious ``overshoots'' in the vicinity of shock fronts.
		These errors compound when integrated in time, such that a naive finite-difference method at this resolution quickly diverges.
		In contrast, the neural network interpolation is so close to the exact solution that it cannot be visually distinguished.
		(b) Histogram of exact vs.\ interpolated function values over our full validation dataset.
		The neural network vastly reduces the number of poor predictions.
		(c) Absolute error vs.\ local curvature.
		The bold line shows the median and shaded region shows the central 90\% of the distribution over the validation set.
		The neural network makes much smaller errors in regions of high curvature, which correspond to shocks.
	}
	\label{fig:regression}
\end{figure}

With this in mind, consider a typical segment of a solution to Burgers' equation (Fig.~\ref{fig:regression}(a)).
We would like to compute the time derivative of the field given a low-resolution set of points (blue diamonds in Fig.~\ref{fig:regression}).
Standard finite difference formulas predict this time derivative by approximating $v$ as a piecewise-polynomial function passing through the given points (orange curves in Fig.~\ref{fig:regression}).
But solutions to Burger's equations are not polynomials, they are shocks with characteristic properties.
By using this information, we can derive a more accurate, albeit equation specific, formula for the spatial derivatives.
For the method to work it should be possible to reconstruct the fine-scale solution from low resolution data.
To this end, we ran many simulations of \eqref{eq:burgers} and used the resulting data to train a neural network.
Fig.~\ref{fig:regression} compares the predictions of our neural net (details below and in SI Appendix) to 4th order polynomial interpolation.
This learned model is clearly far superior to the polynomial approximation, demonstrating that the spatial resolution required for parameterizing the solution manifold can be greatly reduced with equation-specific approximations rather than finite differences.

\section{Models for time integration}

The natural question to ask next is whether such parameterizations can be used for time integration.
For this to work well, integration in time must be numerically stable, and our models need a strong generalization capacity: even a single error could throw off the solution for later times.

To achieve this, we use multi-layer neural networks to parametrize the solution manifold, because of their flexibility, including the ability to impose physical constraints and interpretability through choice of model architecture.
The high-level aspects of the network's design, which we believe are of general interest, are described below.
Additional technical details are described in the SI Appendix and source code is available online at \url{https://github.com/google/data-driven-discretization-1d}.

\paragraph*{Pseudo-linear representation}
Our network represents spatial derivatives with a generalized finite-difference formula similar to \eqref{eq:FD}: the output of the network is a list of coefficients $\alpha_1,\dots,\alpha_N$ such that the $n$-th derivative is expressed as a pseudo-linear filter, \eqref{eq:FD}, where the coefficients $\alpha_i^{(n)}(v_{1},v_2,\dots)$ depend on space and time through their dependence on the field values in the neighboring cells. Finding the optimal coefficients is the crux of our method.

The pseudo-linear representation is a direct generalization of the finite-difference scheme of~\eqref{eq:FD}.
Moreover, exactly as in the case of \eqref{eq:FD}, a Taylor expansion allows us to guarantee formal polynomial accuracy.
That is, we can impose that approximation errors decay as $\mathcal{O}(\Delta x^m)$ for some $m\le N - n$, by layering a fixed affine transformation (see SI).
We found the best results when imposing linear accuracy, $m=1$ with a 6-point stencil ($N=6$), which we used for all results shown here.
Lastly, we note that this pseudo-linear form is also a generalization of the popular ENO and WENO methods~\cite{Harten1987, WENO_lecture_notes}, which choose a local linear filter (or a combination of filters) from a precomputed list according to an estimate of the solution's local curvature.
WENO is an efficient, human-understandable, way of adaptively choosing filters, inspired by nonlinear approximation theory.
We improve on WENO by replacing heuristics with directly optimized quantities.

\paragraph*{Physical constraints}
Since Burgers' equation is an instance of the continuity equation, as with traditional methods, a major increase in stability is obtained when using a finite-volume scheme, ensuring the coarse-grained solution satisfies the conservation law implied by the continuity equation.
That is, coarse-grained equations are derived for the cell averages of the field $v$, rather than its nodal values~\cite{Leveque}.
During training we provide the cell average to the network as the ``true'' value of the discretized field.

Integrating \eqref{eq:burgers}, it is seen that the change rate of the cell averages  is completely determined by the fluxes at cell boundaries.
This is an exact relation, in which the only challenge is estimating the flux given the cell averages.
Thus, prediction is carried out in three steps: first, the network reconstructs the spatial derivatives on the boundary between grid cells (staggered grid).
Then, the approximated derivatives are used to calculate the flux $J$ using the exact formula \eqref{eq:burgers}.
Lastly, the temporal derivative of the cell averages is obtained by calculating the total change at each cell by subtracting $J$ at the cell's left and right boundaries.
The calculation of the time derivative from the flux can also be done using traditional techniques that promote stability, such as monotone numerical fluxes~\cite{Leveque}.
For some experiments, we use Godunov flux, inspired by finite-volume ENO schemes~\cite{Harten1987, WENO_lecture_notes}, but it did not improve predictions for our neural networks models.

Dividing the inference procedure into these steps is favorable in a few aspects:
First, it allows to constrain the model at the various stages using traditional techniques: the conservative constraint, numerical flux and formal polynomial accuracy constraints are what we use here, but other constraints are also conceivable.
Second, this scheme limits the machine-learning part to reconstructing the unknown solution at cell boundaries, which is the main conceptual challenge, while the rest of the scheme follows either the exact dynamics or traditional approximations for it.
Third, it makes the trained model more interpretable since the intermediate outputs (e.g., $J$ or $\alpha_i$) have clear physical meaning.
Lastly, these physical constraints contribute to more accurate and stable models, as detailed in the ablation study in the SI Appendix.

\paragraph*{Choice of loss}

The loss of a neural net is the objective function minimized during training.
Rather than optimizing the prediction accuracy of the spatial derivatives, we optimize the accuracy of the resulting time derivative\footnote{For one specific case, namely the constant-coefficient model of Burgers' Equation with Godunov flux limiting, trained models showed poor performance (e.g., not monotonically increasing with resample factor) unless the loss explicitly included the time-integrated solution, as done in~\cite{Tompson2017}. Results shown in Figs.~\ref{fig:mainresults} \& \ref{fig:survival_plot} use this loss for the constant coefficient models with Burgers' Equation. See details in the SI.}.
This allows us to incorporate physical constraints in the training procedure and directly optimize the final predictions rather than intermediate stages.
Our loss is the mean squared error between the predicted time derivative and labeled data produced by coarse graining the fully-resolved simulations.

Note that a low value of our training loss is a necessary but not sufficient condition for accurate and stable numerical integration over time.
Many models with low training loss exhibited poor stability when numerically integrated (e.g., without the conservative constraint), particularly for equations with low dissipation.
From a machine learning perspective, this is   unsurprising: imitation learning approaches, such as our models, often exhibit such issues because the distribution of inputs produced by the model's own predictions can differ from the training data~\cite{ross2010}.
Incorporating the time-integrated solution into the loss improved predictions in some cases (as in~\cite{Tompson2017}), but did not guarantee stability, and could cause the training procedure itself to diverge due to decreased stability in calculating the loss.
Stability for learned numerical methods remains an important area of exploration for future work.

\paragraph*{Learned coefficients}

We consider two different parameterizations for learned coefficients.
In our first parametrization, we learn optimized time- and space-independent coefficients.
These fixed coefficients minimize the loss when averaged over the whole training set for a particular equation, without allowing the scheme to adapt the coefficients according to local features of the solution.
Below, we refer to these as ``optimized constant coefficients''.
In our second parametrization, we allow the coefficients to be an arbitrary function of the neighboring field values $\{v_i\}$, implemented as a fully-convolutional neural network~\cite{Goodfellow-et-al-2016}.
We use the exact same architecture (three layers, each with 32 filters, kernel size of 5 and ReLU nonlinearity) for coarse-graining all equations discussed in this work.

\begin{figure}
	\centering
	\includegraphics[width=3.42in]{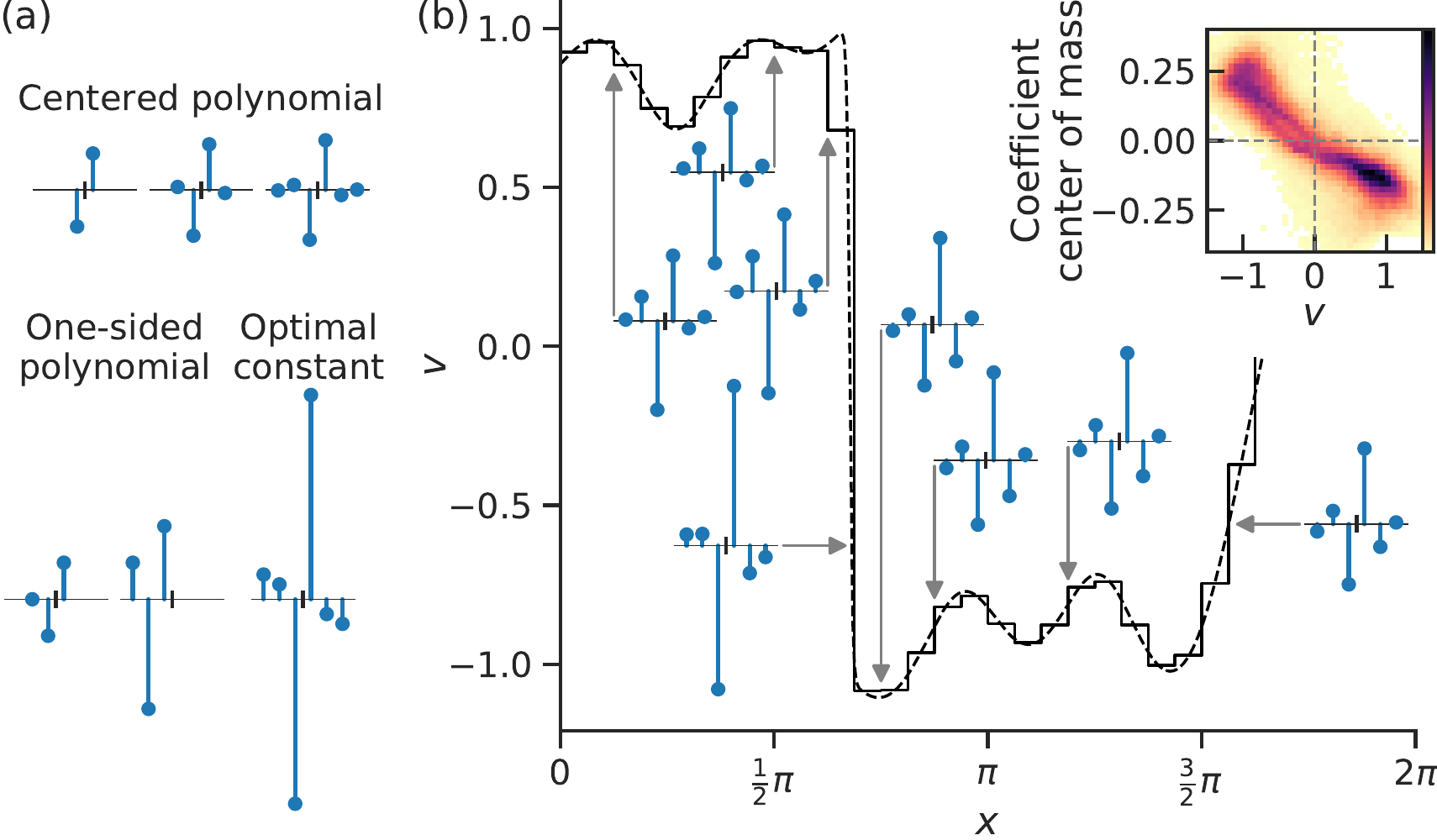}
	\caption{
		\textbf{Learned finite volume coefficients for Burgers' equation.}
		Fixed and spatiotemporally varying finite volume coefficients $\alpha_1^{(1)},\dots,\alpha_6^{(1)}$ (see \eqref{eq:FD}) for $\partial v / \partial x$.
		(a) Various centered and one-sided polynomial finite volume coefficients, along with optimized constant coefficients trained on this dataset (16x resample factor in Figure~\ref{fig:mainresults}).
		The vertical scale, which is the same for all coefficient plots, is not shown for clarity.
		(b) An example temporal snapshot of a solution to Burgers' equation [\eqref{eq:burgers}], along with data-dependent coefficients produced by our neural network model at each of the indicated positions on cell boundaries.
		The continuous solution is plotted as a dashed line, and the discrete cell-averaged representation is plotted as a piecewise constant solid line.
		The optimized constant coefficients are most similar to the neural network's coefficients at the shock position.
		Away from the shock, the solution resembles centered polynomial coefficients.
		(Inset) Relative probability density for neural network coefficient ``center of mass'' vs.\ field value $v$ across our full test dataset.
		Center of mass is calculated by averaging the positions of each element in the stencil, weighted by the absolute value of the coefficient.
	}
	\label{fig:burgers-coefficients}
\end{figure}

Example coefficients predicted by our trained models are shown in Fig.~\ref{fig:burgers-coefficients} and SI Appendix, Fig.~S3.
Both the optimized constant and data-dependent coefficients differ from baseline polynomial schemes, particularly in the vicinity of the shock.
The neural network solutions are particularly interesting: they do not appear to be using one-sided stencils near the shock, in contrast to traditional numerical methods such as WENO~\cite{WENO_lecture_notes} which avoid placing large weights across discontinuities.

The output coefficients can also be interpreted physically.
For example, coefficients for both $\partial v / \partial x$ (Fig.~\ref{fig:burgers-coefficients}(b) inset) and $v$ (SI Appendix, Fig.~S3c) are  either right- or left-biased, opposite the sign of $v$.
This is in line  with our physical intuition: Burgers' equation describes fluid flow, and the sign of $v$ corresponds to the direction of flow.
Coefficients that are biased in the opposite direction of $v$ essentially look ``upwind,'' a standard strategy in traditional numerical methods for solving hyperbolic PDEs~\cite{Leveque}, which helps constrain the scheme from violating temporal causality.
Alternatively, upwinding could be built into the model structure by construction, as we do in models which use Godunov flux.

\section{Results}

\paragraph*{Burgers' equation}
To assess the accuracy of the time integration from our coarse grained model, we computed ``exact'' solutions to \eqref{eq:burgers} for different realizations of $f(x,t)$ at high enough resolution to ensure mesh convergence.
These realizations of $f$ were drawn from the same distribution as the those used for training, but were not in the training set.
Then, for the same realization of the forcing we solved the equation at a lower, possibly under-resolved resolution using four different methods for calculating the flux: (a) a standard finite volume scheme with either 1st order or 3rd order accuracy;
(b) 5th order upwind-biased WENO scheme with Godunov flux~\cite{WENO_lecture_notes};
(c) spatial derivatives estimated by constant optimized coefficients, with and without Godonov flux; and
(d) spatial derivatives estimated by the space- and time-dependent coefficients, computed with a neural net.

\begin{figure}
	\centering
	\includegraphics[width=3.42in]{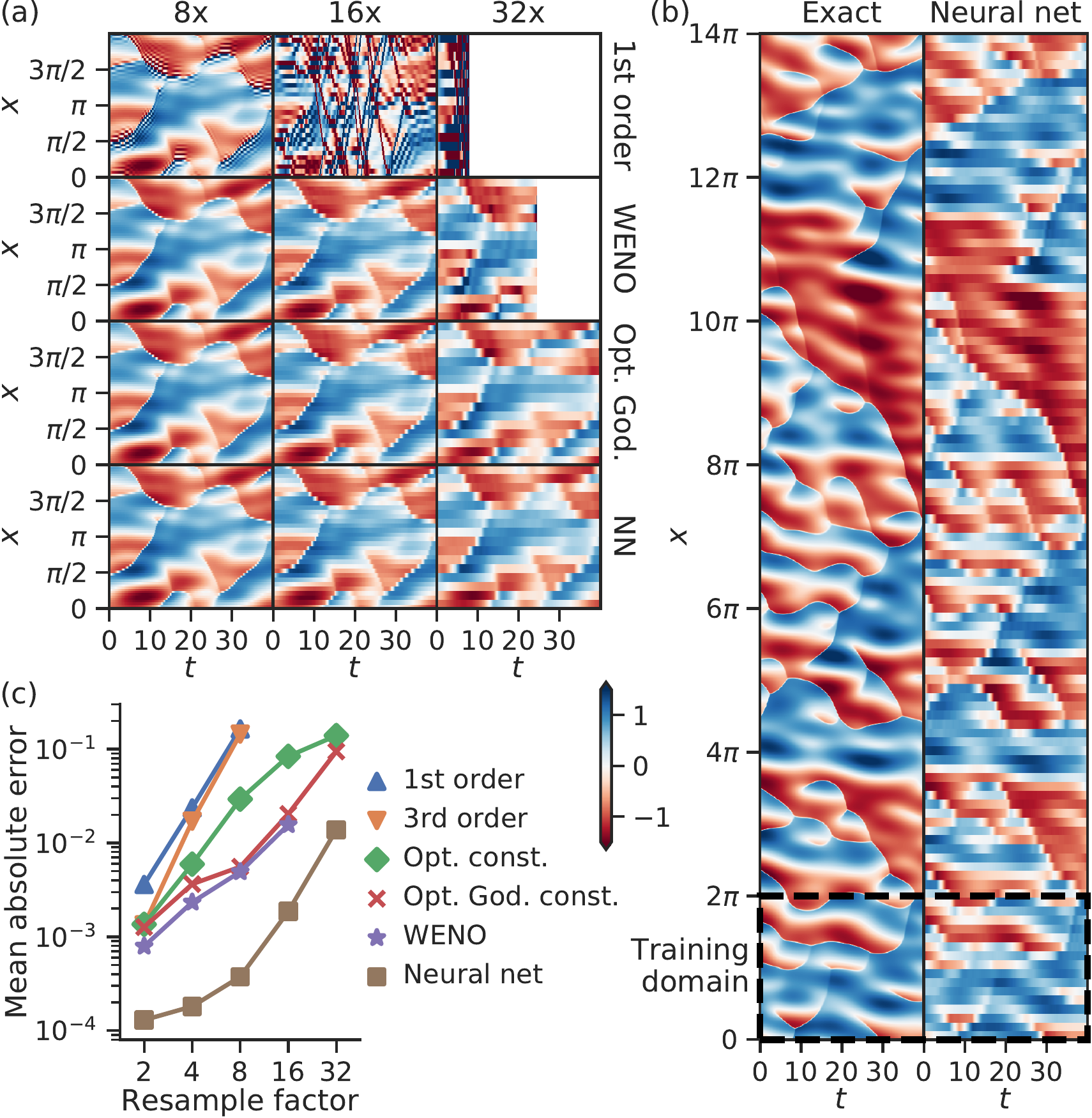}
	\caption{\textbf{Time integration results for Burgers' equation.}
		(a) A particular realization of a solution at varying resolution solved by the baseline 1st order finite volume method, WENO, optimized constant coefficients with Godunov flux, and the neural network, with the white region indicating times when the solution diverged.
		Both learned methods manifestly outperform the baseline method, and even outperform WENO at coarse resolutions.
		(b) Inference predictions for the 32x neural network model, on a ten times larger spatial domain (only partially shown).
		The box surrounded by the dashed line shows the spatial domain used for training.
		(c) Mean absolute error between integrated solutions and the ground truth, averaged over space, times less than 15, and 10 forcing realizations on the ten-times larger inference domain.
		These metrics almost exactly match results on the smaller training domain $[0, 2\pi]$ (Fig.~\ref{fig:mae-for-all-1x-and-10x}).
		As ground truth, we use WENO simulations on a 1x grid.
		Markers are omitted if some simulations diverged or if the average error is worse than fixing $v=0$.
	}
	\label{fig:mainresults}
\end{figure}
Results are shown in Fig.~\ref{fig:mainresults}.
Panel (a) compares the integration results for a particular realization of the forcing for different values of the \textit{resample factor}, that is, the ratio between the number of grid points in the low resolution calculation and that of the fully converged solution\footnote{Physically, the natural measure of the spatial resolution is with respect to the internal length-scale of the equation which in the case of Burgers' equation is the typical shock width.
	However, since this analysis is meant to be applicable also to situations where the internal length-scale is a-priori unknown, we compare here to the length-scale at which mesh convergence is obtained. }.
Our learned models, both with constant and solution-dependent coefficients, can propagate the solution in time and dramatically outperform the baseline method at low resolution.
Importantly, the ringing effect around the shocks, which leads to numerical instabilities, is practically eliminated.

Since our model is trained on fully resolved simulations, a crucial requirement for our method to be of practical use is that training can be done on small systems, but still produce models that perform well on larger ones.
We expect this to be the case, since our models, being based on convolutional neural networks, use only local features and by construction are translation invariant.
Panel (b) illustrates the performance of our model trained on the domain $[0, 2\pi]$ for predictions on a ten times larger spatial domain of size $[0, 20\pi]$.
The learned model generalizes well.
For example, it shows good performance by when function values are all positive in a region of size greater than $2\pi$, which due to the conservation law cannot occur in the training dataset.

\begin{figure}
	\centering
	\includegraphics[width=3.42in]{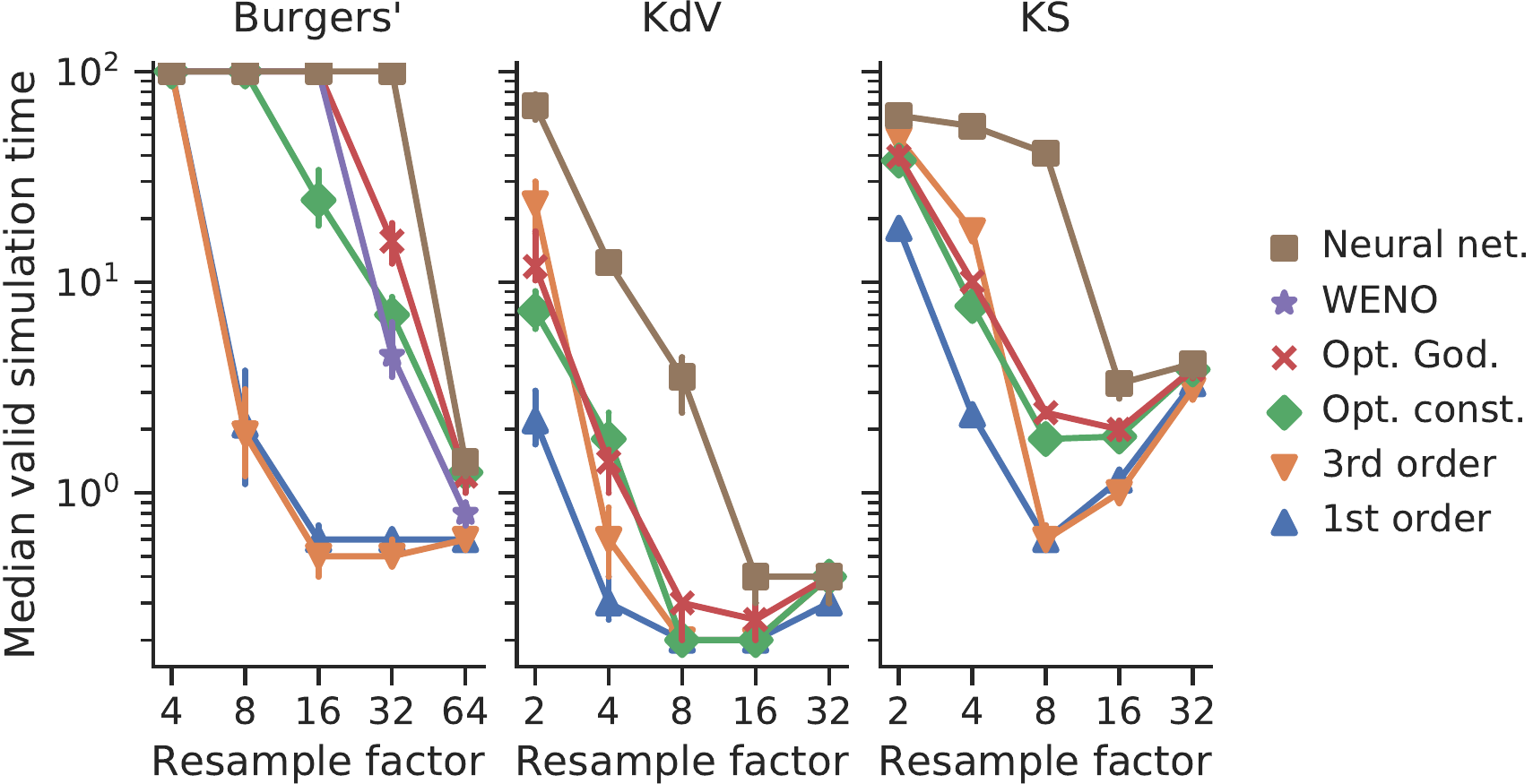}
	\caption{
		\textbf{Model performance across all of our test equations.}
		Each plot shows the median time for which an integrated solution remains ``valid'' for each equation, defined by the absolute error on at least 80\% of grid points being less than the 20th percentile of the absolute error from predicting all zeros.
		These thresholds were chosen so that ``valid'' corresponds to relatively generous definition of an approximately correct solution.
		Error bars show the 95\% confidence interval for the median across 100 simulations for each equation, determined by bootstrap resampling.
		Simulations for each equation were run out to a maximum of time of 100.
	}
	\label{fig:survival_plot}
\end{figure}

To make this assessment quantitative, we averaged over many realizations of the forcing and calculated the mean absolute error integrated over time and space.
Results on the ten-times larger inference domain are shown in panel (c): the solution from the full neural network has equivalent accuracy to increasing the resolution for the baseline by a factor of about 8x.
Interestingly, even the simpler constant-coefficient method significantly outperforms the baseline scheme.
The constant coefficient model with Godunov flux is particularly compelling.
This model is faster than WENO, because there is no need to calculate coefficients on the fly, with comparable accuracy and better numerical stability at coarse resolution, as shown in panel (a) and Fig.~\ref{fig:survival_plot}.

These calculations demonstrate that neural networks can carry out coarse graining.
Even if the mesh spacing is much larger than the shock width, the model is still able to accurately propagate dynamics over time, showing that it has learned an internal representation of the shock structure.

\paragraph*{Other examples}

To demonstrate the robustness of this method, we repeated the procedure for two other canonical PDEs:
the Korteweg-de Vries (KdV) equation~\cite{Zabusky1965}, which was first derived to model solitary waves on a river bore and is known for being completely integrable and to feature soliton solutions; and the Kuramoto-Sivashinsky (KS) equation which models flame fronts and is a textbook example of a classically chaotic PDE~\cite{DE_handbook}.
All details about these equations are given in the SI.
We repeated the training procedure outlined above for these equations, running high resolution simulations and collecting data to train equation-specific estimators of the spatial derivative based on a coarse grid.
These equations are essentially non-dissipative, so we do not include a forcing term.
The solution manifold is explored by changing the initial conditions, which are taken to be a superposition of long-wavelength sinusoidal functions with random amplitudes and phases (see SI for details).

To assess the accuracy of the integrated solution, for each initial condition we define ``valid simulation time'' as the first time that the low-resolution integrated solution deviates from the cell-averaged high-resolution solution by more than a given threshold.
We found this metric more informative to compare across very different equations than absolute error.

Figure~\ref{fig:survival_plot} shows the median valid simulation time as a function of the resample factor.
For all equations and resolutions, our neural network models have comparable or better performance than all other methods.
The neural network is particularly advantageous at low resolutions, demonstrating its improved ability to solve coarse-grained dynamics.
The optimized constant coefficients perform better at coarse resolution than baseline methods, but not always at high resolutions.
Finally, at large enough resample factors the neural network approximations also fails to reproduce the dynamics, as expected.
These results also hold on a ten-times larger spatial domain, as shown in the SI, along with figures illustrating specific realizations and mean absolute error.

\section{Discussion and conclusion}

It has long been remarked that even simple nonlinear PDEs can generate solutions of great complexity.
But even very complex, possibly chaotic, solutions are not just arbitrary functions: they are highly constrained by the equations they solve.
In mathematical terms, despite the fact that the solution set of a PDE is nominally infinite dimensional, the inertial manifold of solutions is much smaller, and can be understood in terms of interactions between local features of the solutions to nonlinear PDEs.
The dynamical rules for interactions between these features have been well studied over the past fifty years.
Examples include, among many others, interactions of shocks in complex media, interactions of solitons~\cite{Zabusky1965}, and the turbulent energy cascade~\cite{Frisch1996}.

Machine learning offers a different approach for modeling these phenomena, by using training data to parametrize the inertial manifold itself; said differently, it learns both the features and their interactions from experience of the solutions.
Here we propose a simple algorithm for achieving this, motivated by coarse graining in physical systems.
It is often the case that coarse graining a PDE amounts to modifying the weights in a discretized numerical scheme.
Instead, we use known solutions to learn these weights directly, generating \emph{data driven discretizations}.
This effectively parametrizes the solution manifold of the PDE, allowing the equation to be solved at high accuracy with an unprecedented low resolution.

Faced with this success, it is tempting to try and leverage the understanding the neural network has developed in order to gain new insights about the equation or its coarse-grained representation.
Indeed, in Fig.~\ref{fig:burgers-coefficients} we could clearly interpret the directionality of the weights as an upwind bias, the pseudolinear representation providing a clear interpretation of the prediction in a physically sensible way.
However, extracting more abstract insight from network, such as the scaling relation between the shock height and width is a difficult challenge.
This is a general problem in the field of machine learning, which is under intensive current research~\cite{Sundararajan2017, Shrikumar2017}.

Our results are promising, but two challenges remain before our approach can be deployed at large scales.
The first challenge is speed.
We showed that optimized constant coefficients can already improve accuracy, but our best models rely on the flexibility of neural networks.
Unfortunately, our neural nets use many more convolution operations than the single convolution required to implement finite differences, e.g., $32^2 = 1024$ convolutions with a five point stencil between our second and third layers.
We suspect that other machine learning approaches could be dramatically faster.
For example, recent work on a related problem -- inferring sub-pixel resolution from natural images -- has shown that banks of pretrained linear filters can nearly match the accuracy of neural nets with orders of magnitude better performance~\cite{Romano2016, Getreuer2018blade}.
The basic idea is to divide input images into local patches, classify patches into classes based on fixed properties (e.g., curvature and orientation), and learn a single optimal linear filter for each class.
Such computational architectures would also facilitate extracting physical insights from trained filters.

A second challenge is scaling to higher dimensional problems and more complex grids.
Here we showcased the approach for regular grids in one dimension, but most problems in the real world are higher dimensional, and irregular and adaptive grids are common.
We do expect larger potential gains in two and three dimensions, as the computational gain in terms of the number of grid points would scale like the square or the cube of the resample factor.
Irregular grids may be more challenging, but deep learning methods that respect appropriate invariants have been developed both for arbitrary graphs~\cite{Gilmer2017} and collections of points in 3D space~\cite{Qi2017}.
Similar to what we found here, we expect that hand-tuned heuristics for both gridding and grid coefficients could be improved upon by systematic machine learning.
More broadly, data driven discretization suggests the potential of data driven numerical methods, combining the optimized approximations of machine learning with the generalization of physical laws.

\paragraph*{Acknowledgements}
We thank Peyman Milanfar, Pascal Getreur, Ignacio Garcia Dorado and Dmitrii Kochkov for collaboration and important conversations, Peter Norgaard and Geoff Davis for feedback on drafts of the manuscript, and Chi-Wang Shu for guidance on the implementation of WENO.
Y.B.S acknowledges support from the James S.~McDonnel post-doctoral fellowship for the study of complex systems.
M.P.B acknowledges support from NSF DMS-1715477 as well as the Simons Foundation.

\newpage
\renewcommand*{\bibfont}{\small}

\clearpage

\begin{center}
\textbf{\large Supplemental Materials}
\vspace{1cm}
\end{center}

\renewcommand{\thefigure}{S\arabic{figure}}
\renewcommand{\theequation}{S-\arabic{equation}}
\setcounter{figure}{0}
\setcounter{equation}{0}
\begin{figure*}[tbh]
	\centering
	\includegraphics[width=0.8\linewidth]{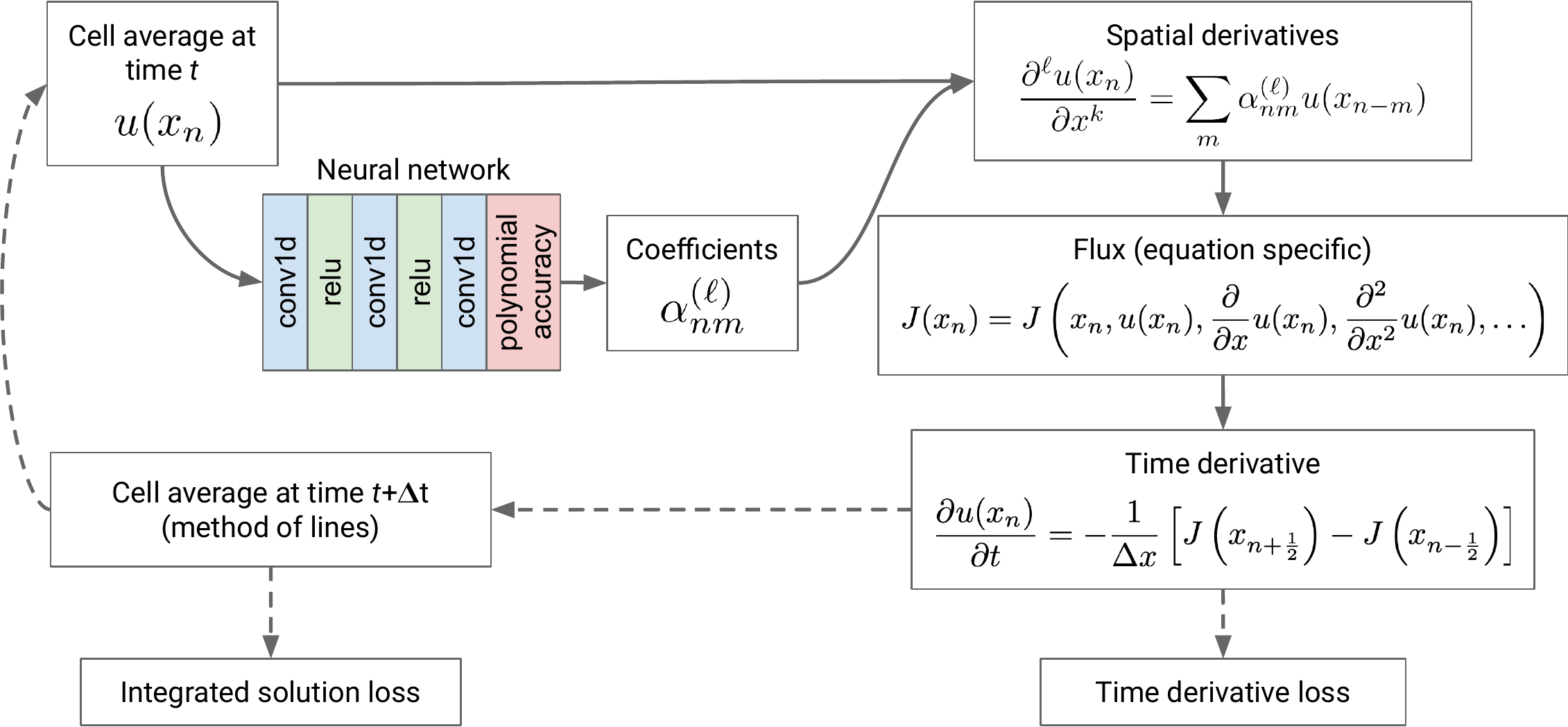}
	\caption{
		\textbf{Neural network architecture.}
		During training, the model is optimized to predict cell average time-derivatives or time evolved solution values from cell average values, based on a precomputed dataset of snapshots from high resolution simulations.
		During inference, the optimized model is repeatedly applied to predict time evolution using the method of lines.
	}
	\label{fig:neural-net-architecture}
\end{figure*}

\section*{Appendix I: Neural network model}

\paragraph*{Model details}

Complete source code, allowing production of a training dataset, training the network on it, and deployment of the resulting coarse-grained equation is freely available online at \url{https://github.com/google/data-driven-discretization-1d}.
Our model, including all physical constraints, was implemented using the TensorFlow library~\cite{tensorflow}.
In the calculations presented here, the model had three fully convolutional layers, each with 32 filters of a fixed kernel of size five and with a ReLU nonlinearity between each layer.
Our neural network predictions at a single point are thus dependent on values of the local solution over a maximum range of 13 grid cells, independent of the model resolution.
The code allows for easily tuning these hyper-parameters.

Fig.~\ref{fig:neural-net-architecture} presents a graphical depiction of our model architecture.
The coarse grained function values are fed into a neural network.
The network's output, the coefficients $\alpha_{i}^{(n)}$, are combined with the coarse grained values to estimate the spatial derivatives.
These are used in the known physical equation for the flux, which is used to calculate the temporal derivative by a first-order divergence.
Training minimizes either the difference between the calculated time derivative and the true one (most models), or the difference between the calculated evolved state at future times and the true evolved state (Burgers' equation with constant coefficients only, as noted below).

We trained our models using the Adam optimizer for \num{40000} steps total, decreasing the learning rate by a factor of 10 after \num{20000} steps.
For most models we used an initial learning rate of \num{3e-3}, with the exception of KdV and KS models with a resample factor of 16x and higher, for which we used an initial learning rate of \num{1e-3}.
We used a batch size of 128 times the resampling factor.
All of our results show models trained with the time-derivative loss, with the exception of the ``optimized constant coefficient'' models for Burgers' equation, which was trained to predict 8 forward time-steps with the midpoint method.
Each individual model trained to completion in less than an hour on a single Nvidia P100 GPU.

We found that some of our models had highly variable performance on different model training runs, due to randomness in the training procedure and a lack of guarantees of numerical stability in our training procedure.
To ameliorate this issue and improve the interpretability of our results, we trained each model ten times and in most results only show predictions from only the best overall performing model for each task.
Predictions from the worst-performing models are shown below in~\ref{fig:ablation}.

\paragraph*{Godunov numerical flux}

For some models, we used Godunov numerical flux for the convective term $v^2$ in the flux.
Following the example of numerical fluxes for WENO methods~\cite{WENO_lecture_notes}, we construct both left- and right-sided estimates of $v$ (with separate $\alpha$ coefficients for each), $v^-$ and $v^+$, and combine them according to the Godunov flux rule for $J(v) = v^2$:
\begin{align}
J_\text{godunov}(v^-, v^+) &=
\begin{cases}
\min[(v^-)^2, (v^+)^2] & \text{if $v^- \leq v^+$} \\
\max[(v^-)^2, (v^+)^2] & \text{if $v^- > v^+$}
\end{cases}.
\label{eq:godunov}
\end{align}

\section*{Appendix II: PDE parameters}

\paragraph*{Equations}
We solved three PDEs in one space dimension $x$ and time dimension $t$.
All of our equations can be written in the same conservative form,
\begin{equation}
\pd{v}{t} + \pd{J}{x} = F(x,t)\ ,
\quad
v(x,t=0) = v_0(x),
\end{equation}
with different choices for the flux $J$, forcing $F$ and initial conditions $v_0$:
\begin{widetext}
\begin{alignat}{4}
\mbox{Burgers:}& &\qquad J \equiv & \frac{v^2}{2} - \eta \pd{v}{x}, &\qquad F \equiv & f(x, t), &\qquad v_0 \equiv & 0 \\
\mbox{Kortweg-de Vries (KdV):}& & J \equiv & 3 v^2 + \pdd{v}{x}, & F \equiv & 0, & v_0 \equiv & f(x, 0) \\
\mbox{Kuramoto-Shivashinski (KS):}& & J \equiv & \frac{v^2}{2} + \pd{v}{x} + \pddd{v}{x}, & F \equiv& 0, & v_0 \equiv & f(x, 0).
\end{alignat}
\end{widetext}
where the function $f(x, t)$, described below, is initialized with random components to allow for sampling over a manifold of solution.
The form of these equations is written to emphasize their similar structure; the standard form of KdV can be obtained by substituting $v \to -v$.
All equations employ periodic boundary conditions.
For training, we used domains of size $L = 2\pi$ for Burgers' equation, $L = 32$ for KdV and $L=64$ for KS.
For Burgers' equation, we set $\eta = 0.01$.

\paragraph*{Random parameters}

In order to explore the solution manifold, we introduce randomness either to the forcing (Burgers' equation) or to the initial conditions (KS \& KdV). The random signal that we use is a sum of long-wavelength sinusoidal functions:
\begin{align}
f(x,t) &= \sum_{i=1}^N A_i \sin(\omega_i t + 2\pi \ell_i x / L + \phi_i),
\label{eq:forcing}
\end{align}
with the following parameters:
\begin{center}
	\renewcommand{\arraystretch}{2}
	\newcolumntype{C}[1]{>{\centering\let\newline\\\arraybackslash\hspace{0pt}}m{#1}}
	\begin{tabular}{ |C{1.6cm}|C{2.3cm}|C{2.3cm}| }
		\hline
		& Burgers' & KdV and KS \\
		\hline
		$A$ & \multicolumn{2}{c|}{$[-0.5,0.5]$}\\\hline
		$\omega$ & \multicolumn{2}{c|}{$[-0.4,0.4]$}\\\hline
		$\phi$ & \multicolumn{2}{c|}{$[0,2\pi]$}\\\hline
		$\ell$ &  $\{3, 4, 5, 6\}$ & $\{1,2,3\}$\\\hline
		$N$ & 20 & 10 \\\hline
	\end{tabular}
\end{center}

Each parameter is drawn independently and uniformly from its range. For generating Fig.~3(b) and 3(c) of the main text, i.e., deploying the trained model on a 10x larger domain, $\ell$ was allowed to take on integer value that would result in a wavenumber $k = 2\pi\ell/L$ within the range of the original training data, i.e., $\{30, 31, 32, \ldots, 60 \}$.

\begin{figure*}[t]
	\centering
	\includegraphics[width=6in]{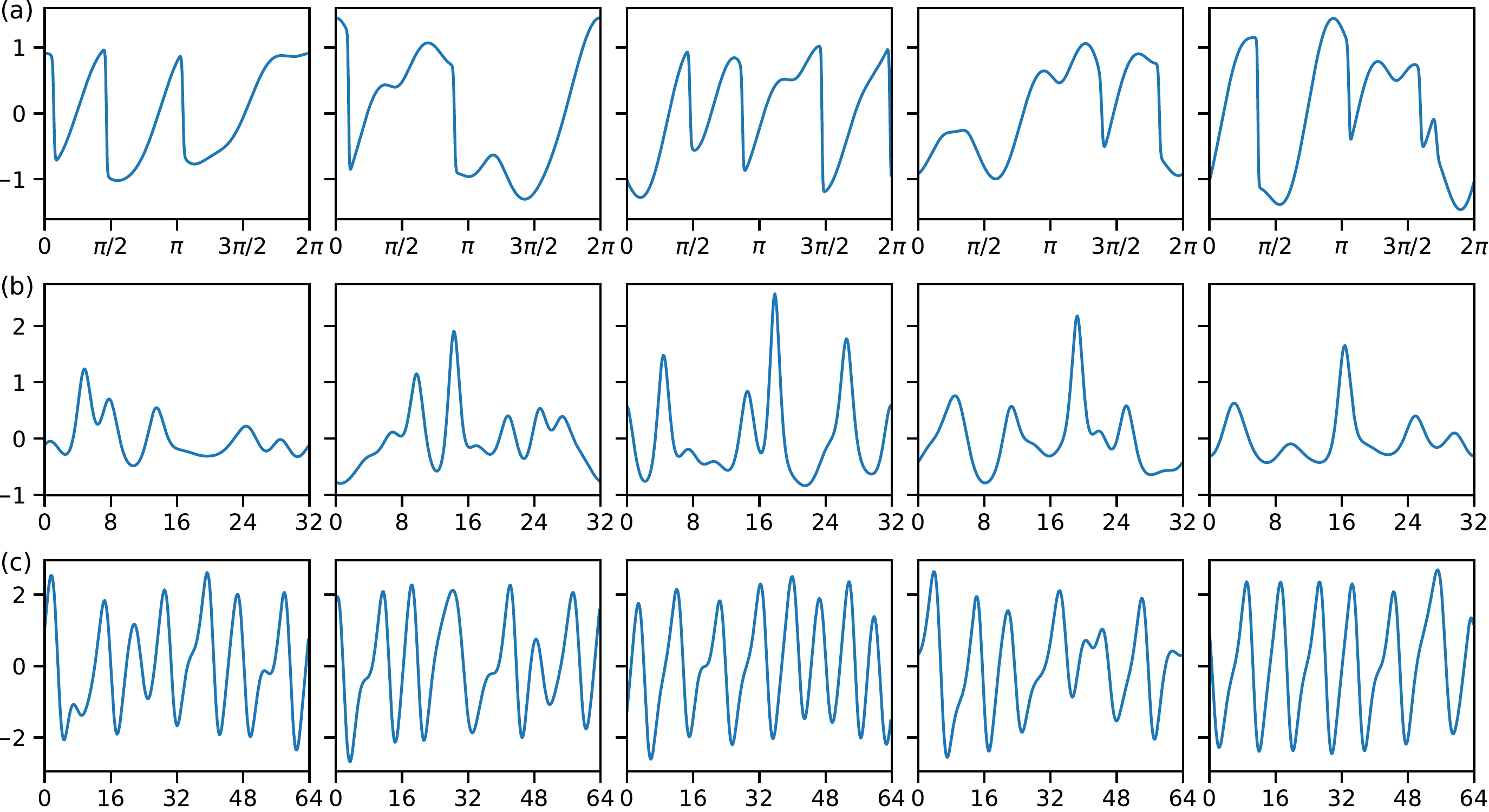}
	\caption{
		Five random samples from the training dataset each for (a) Burgers' equation, (b) KdV and (c) KS.
	}
	\label{fig:training-data-examples}
\end{figure*}

\paragraph*{Training data}
To train the network we generate a set of \si{8000} high-resolution solutions to each equation, sampled at regular time intervals from \si{800} numerical integrations.
Five randomly selected examples for each equation are shown in Fig.~\ref{fig:training-data-examples}
To obtain high accuracy ``exact'' solutions to Burgers' equation on a fully resolved grid, we used a fifth-order WENO method and 512 grids points.
For KdV and KS, we used a pseudo-spectral method for numerical differentiation (which we found to be more accurate than Finite Volumes), with 256 grid points.
However, for calculating the loss during training we used third-order Finite Volumes on the spectral method's solution to calculate spatial derivatives.
For unclear reasons, our learned models for KdV and KS at low resample factors did not perform well when using spectral derivatives for ground-truth.
One possibility is that a lack of access to the global features used by the spectral methods made it impossible for our models (which are spatially local by construction) to exactly copy their behavior.

\paragraph*{Time integration}
For numerical time integration to produce training and evaluation datasets, we used an explicit Runge-Kutta method of order 3(2) with adaptive time-stepping, as implemented in the SciPy package.
For time integration during the training procedure itself, we used the midpoint method with a fixed time-step of 0.001 for Burgers' equation, based on the minimum time-step on the high resolution grid chosen by adaptive time-stepping.
We used these methods due to their robustness and simplicity, but they are certainly \emph{not} an optimized time integration methods for these equations.
For example, a practical method for the viscous Burgers' equation should use implicit time-stepping, because the diffusive term limits the maximum stable explicit time step to be smaller than the square of the grid spacing.
Nonetheless, we believe this is appropriate because the focus of this paper is not on improving the time discretization.

\section*{Appendix III: Polynomial accuracy constraints}
Our method represents the spatial derivative at a point $x_0$ as a linear combination of the function values at $N$ neighbors of $x_0$. That is, for a given derivative order $\ell$, we write
\begin{align}
\left.\frac{\partial^\ell f}{\partial x^\ell}\right|_{x=x_0}=\sum_{n=1}^N \alpha_{n}f(x_0 + h_n)
\label{eq:s1}
\end{align}
where $h_n$ is the offset of the $n$-th neighbor on the coarse grid.
Note that in the main text we only deal with uniformly spaced meshes, $h_n = n\,\Delta x$,  but this formalism holds for an arbitrary mesh.
A crucial advantage of this writing is that we can enforce arbitrary polynomial accuracy up to degree $m$, as long as $m\le N-\ell$, by imposing an affine constraint on the $\alpha$'s.
That is, we can ensure that the error in approximating $f^{(\ell )}$ will be of order $h^{m}$ for some $m\le N-\ell$.

To see this, note  that the standard formula for deriving the finite difference coefficients for the $\ell$-th derivative with an $N$-point stencil is obtained by solving the linear set of equations
\begin{equation}
\begin{pmatrix}
h_1^0 & \cdots & h_N^0 \\
\vdots & \ddots & \vdots \\
h_1^{N-1} & \cdots & h_N^{N-1}
\end{pmatrix}
\begin{pmatrix}
\alpha_{1} \\
\vdots \\
\alpha_N
\end{pmatrix}
=
\ell!
\begin{pmatrix}
\delta_{0,\ell} \\
\vdots\\
\delta_{i,\ell}\\
\vdots\\
\delta_{N-1,\ell}
\end{pmatrix}
\label{eq:fd_matrix}
\end{equation}
where $\delta_{i,j}$ is the Kronecker delta~\cite{fd_formula}.

These equations are obtained by expanding \eqref{eq:s1} in a Taylor series up to order $m-1$ in the vicinity of $x_0$ and requiring equality of order $\O(h^m)$ for arbitrary functions.
Each row in the set of equations corresponds to demanding that a term of order $h^k$ will vanish, for $k=0, 1, \dots, N-1$.
The resulting formula is approximate to polynomial order $N-\ell$~\cite{fd_formula} and the system of equations is fully determined, that is, a unique solution exists, which is obtained by inverting the matrix.
A similar set of linear constraints for finite volume methods can be derived by averaging over each unit cell \cite{WENO_lecture_notes}.

Imposing a lower order approximation amounts to removing the bottom rows from the equation \eqref{eq:fd_matrix}.
Specifically, imposing accuracy of order $m$ amounts to keeping the first $m-N+\ell$ rows, which makes the system under-determined.
Therefore, any solution for $\boldsymbol{\alpha}$ can be written as a sum of an arbitrary fixed solution (say, the standard-finite difference formula of order $m$) plus a vector $\tilde{\boldsymbol{\alpha}}$ from the null-space of the matrix of \eqref{eq:fd_matrix} (with removed rows).

\section*{Appendix IV: Interpolation model}

The neural network model for interpolation of Burgers' equation shown in Fig.~\ref{fig:regression} of the main text uses the same training datasets and a similar model structure to our time integration models.
Instead of predicting cell average values, we train the model to predict the subsampled function values at all intermediate locations from a high-resolution simulation as zeroth order spatial derivatives, imposing first-order accuracy constraints.
We use an 8x subsampled grid, so the outputs of our model are seven interpolated values between each adjacent pair of seed points.
We use the mean squared error at each interpolated point as the loss, scaled such that linear interpolation on the training dataset has a loss of one, and an initial learning rate of \num{1e-3}.
Otherwise, all details match the neural network models described in Appendix I.
The fourth order polynomial interpolation between points $x_i$ and $x_{i+1}$ is based on the function values at $\{v_{i-1}, v_{i}, v_{i+1}, v_{i+1}\}$, which closely corresponds to the interpolation used by traditional finite difference (not finite volume) method.

For Fig.~\ref{fig:regression}(c), we define the curvature of each point as the maximum value of $|\partial^2 v / \partial x^2|$ over the interpolated interval on the high resolution grid.
This heuristic, which is similar to the those used for identifying smooth regions by WENO methods~\cite{WENO_lecture_notes}, provides a good indicator of whether or not the solution is interpolating over a shock front.

\section*{Appendix V: Additional plots of coefficients}

\begin{figure*}
	\centering
	\includegraphics[width=7in]{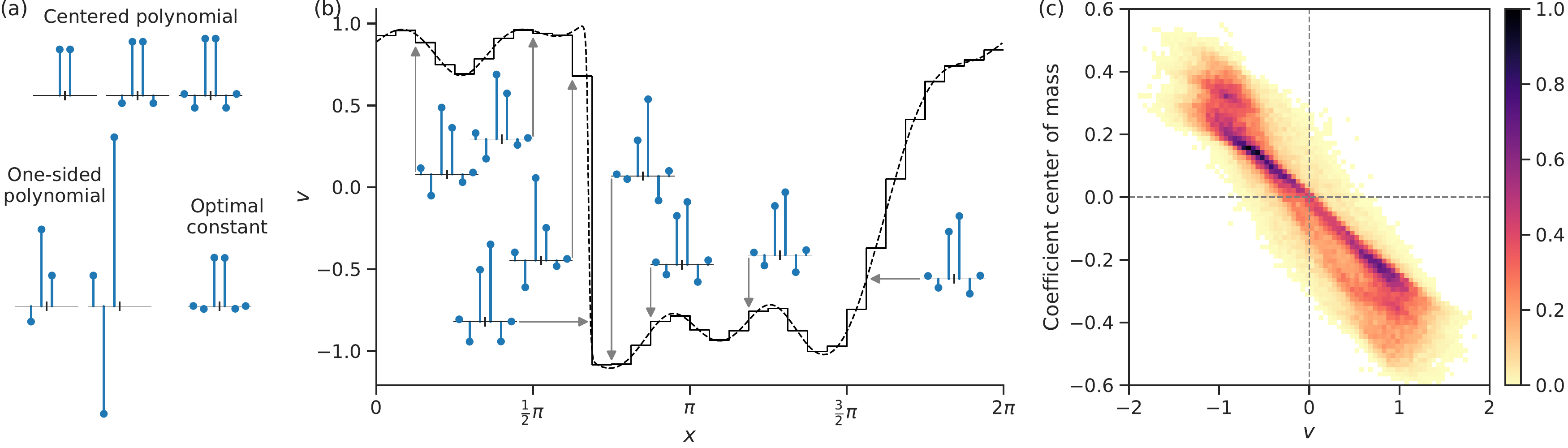}
	\caption{Like Fig.~\ref{fig:burgers-coefficients} of the main text, but showing coefficients for reconstructing the field value $v$ instead of its first derivative.}
	\label{fig:burgers-coefficients-zeroth-derivative}
\end{figure*}

\begin{figure*}
	\centering
	\includegraphics[width=7in]{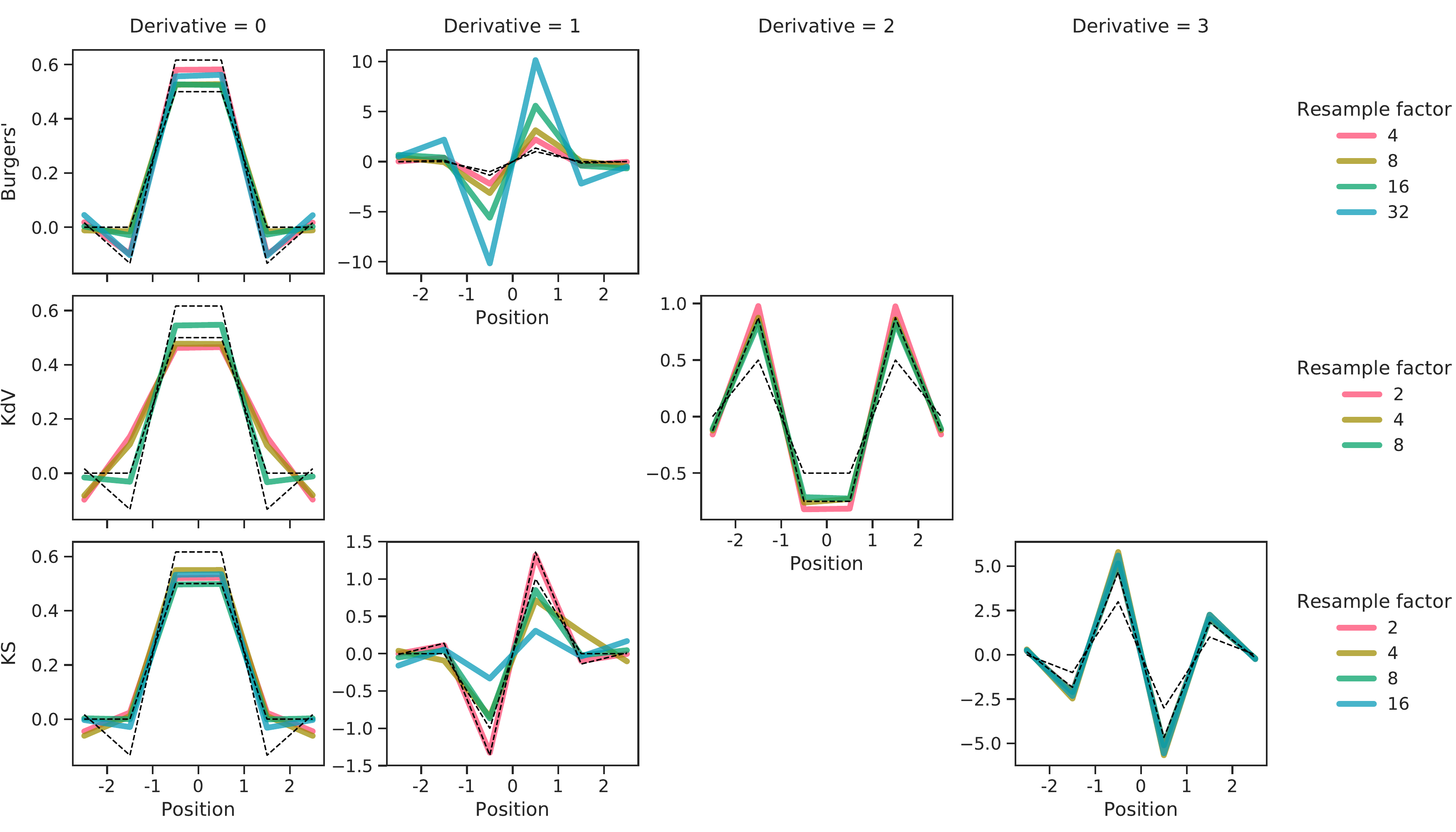}
	\caption{Optimized constant finite volume coefficients for the equations studied in this work at varying resample factor. The dashed black lines correspond to standard centered Finite Volumes coefficients: 2-point and 6-point stencils for zeroth and first derivatives, and 4-point and 6-point stencils for second and third derivatives.}
	\label{fig:optimal-coefficients}
\end{figure*}

\begin{figure*}
	\centering
	\includegraphics[width=7in]{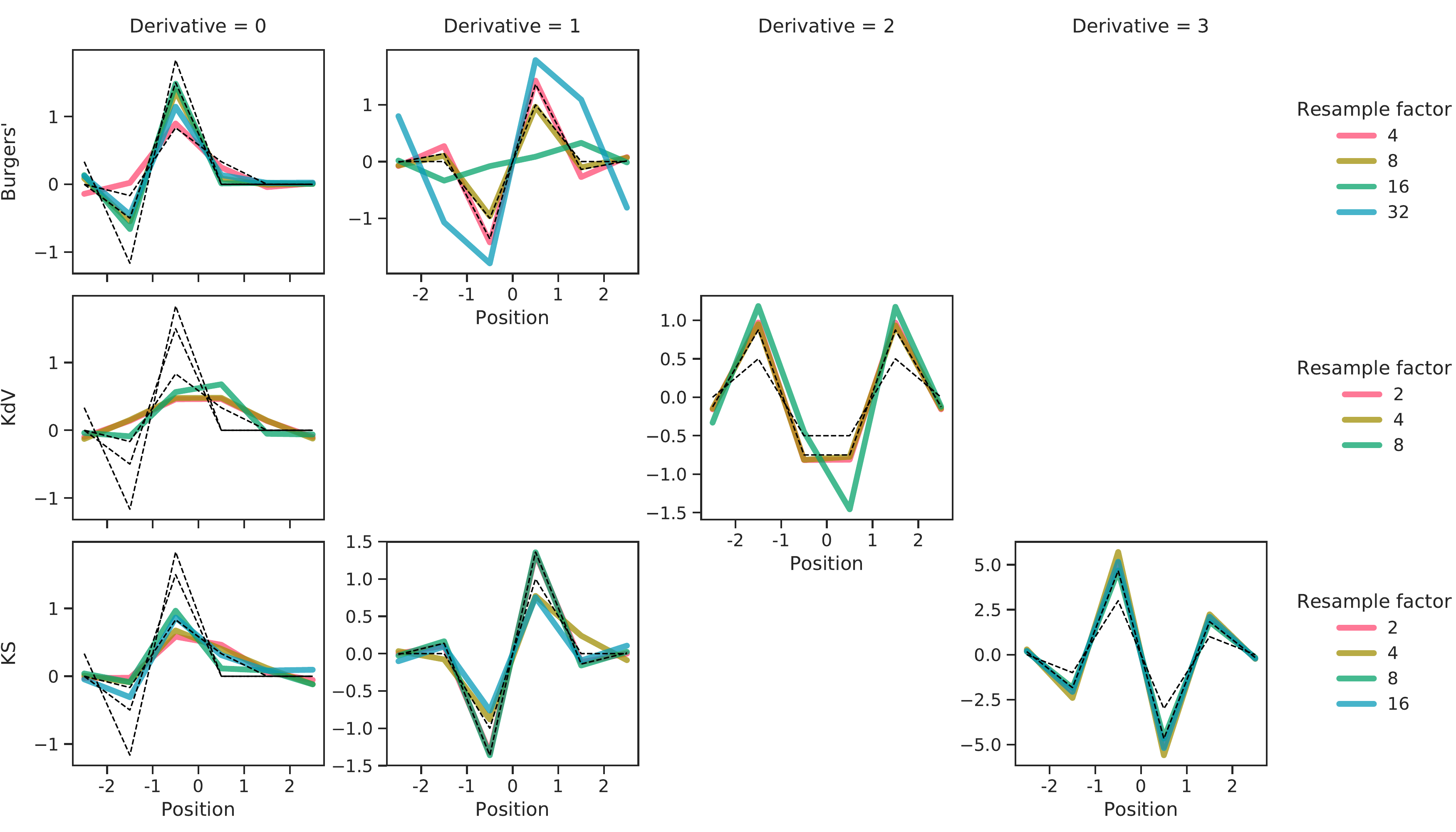}
	\caption{Like Figure~\ref{fig:optimal-coefficients}, but for models that use Godunov monotone flux to combine left- and right-sided estimates $u^{-}$ and $u^{+}$. Only the left-ward facing coefficients $u^{-}$ for reconstructing the zeroth derivative are shown, along standard 2 and 3 point upwinded Finite Volume coefficients.}
	\label{fig:optimal-coefficients-godunov}
\end{figure*}

Figure~\ref{fig:burgers-coefficients-zeroth-derivative} shows learned finite volume coefficients for reconstructing the field value $v$ for Burgers' equation.
Similar to the situation for the coefficients for $\partial v / \partial x$ shown in Fig.~\ref{fig:burgers-coefficients}, the neural network has learned upwinding.

Figure~\ref{fig:optimal-coefficients} shows optimized constant coefficients for Burgers', KdV and KS equations at all resample factors for which the optimized coefficients showed any improvement over standard coefficients.
The optimized coefficients often, but not always, differ significantly from standard centered coefficients, particularly for lower order derivatives.

Figure~\ref{fig:optimal-coefficients-godunov} shows optimized constant coefficients for all equations using the Godunov flux.
For both Burgers' and KS, the optimized coefficients make use of ``upwinding,'' especially at coarser resolutions.

\section*{Appendix VI: Results for KdV and KS equations}

\begin{figure*}
	\centering
	\includegraphics[width=\linewidth]{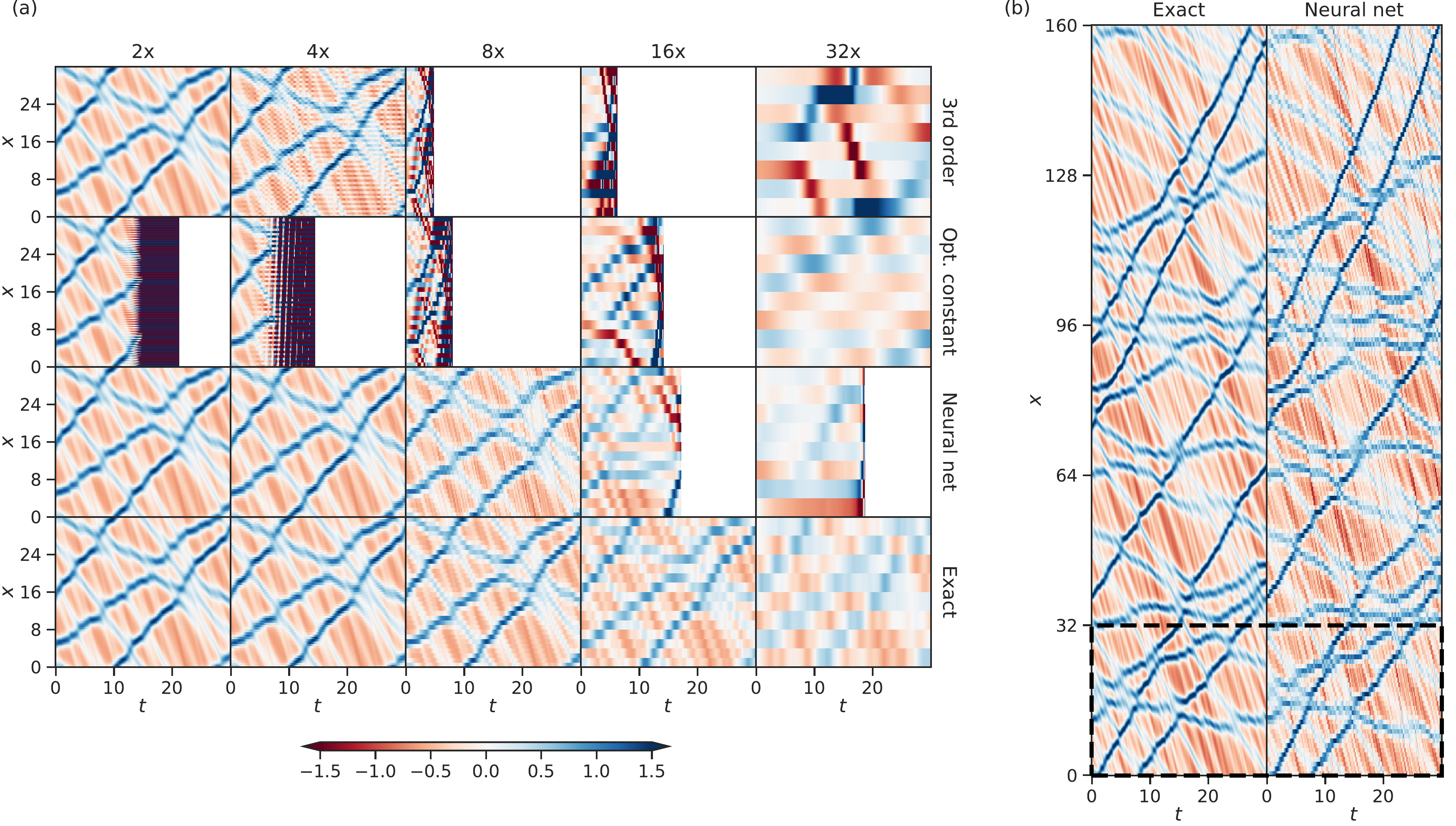}
	\caption{(a) Particular realization of the solution for the Korteweg-de Vries (KdV) equation at varying resolution solved by the baseline 1st order finite volume method (top row), optimal constant coefficients (second row), the neural network (third row) and the exact coarse-grained solution (bottom row). Blank regions indicate where the solver has diverged.
		Note that optimal constant coefficients exhibit lower error at small times than the baseline coefficients, even though the 2x and 4x models suffer from worse numerical stability.
		(b) Inference predictions for the 8x neural network model, on a ten times larger spatial domain (only partially shown). The box surrounded by the dashed line shows the spatial extent of the training domain.
	}
	\label{fig:kdv-example}
\end{figure*}

\begin{figure*}
	\centering
	\includegraphics[width=\linewidth]{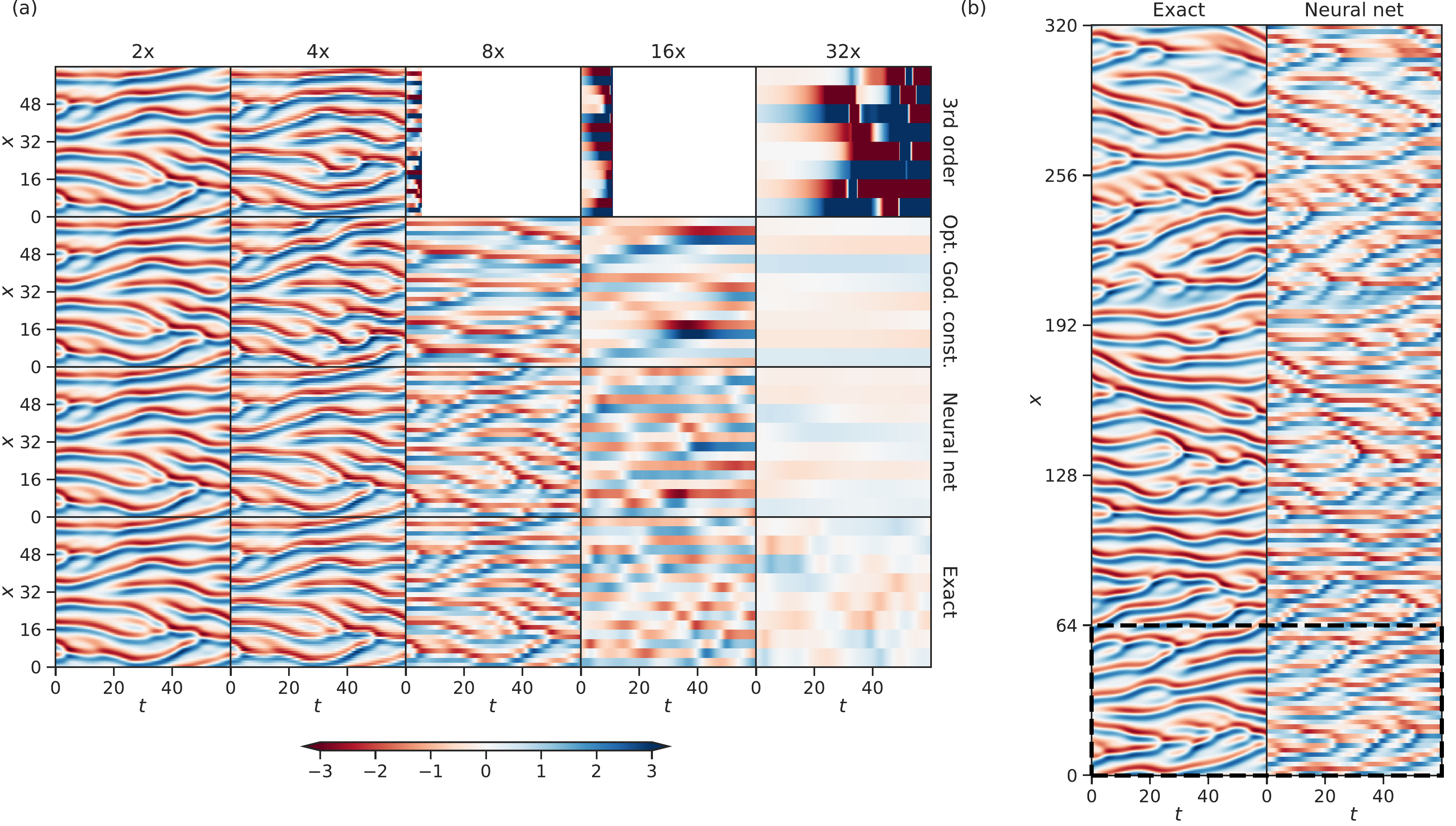}
	\caption{Same as \ref{fig:kdv-example}, but for the Kuramoto-Shivashinsky (KS) equation.}
	\label{fig:ks-example}
\end{figure*}

Similar to Fig.~\ref{fig:mainresults}(a) and \ref{fig:mainresults}(b) in the main text, Figures \ref{fig:kdv-example} and \ref{fig:ks-example} show particular realizations of solutions to the KdV and KS Equations, respectively, on both the a domain of the same size as the training domain and on a 10x larger validation domain (for the 8x resample factor model).
The figures show how the same initial condition is solved at different resample factors by the baseline finite-difference method and neural network, demonstrating that our method significantly outperforms the baseline.
They also show that our models for KdV and KS also generalize to a much larger spatial domain than used for training.

\begin{figure*}[tbh]
    \centering
    \includegraphics[width=6.5in]{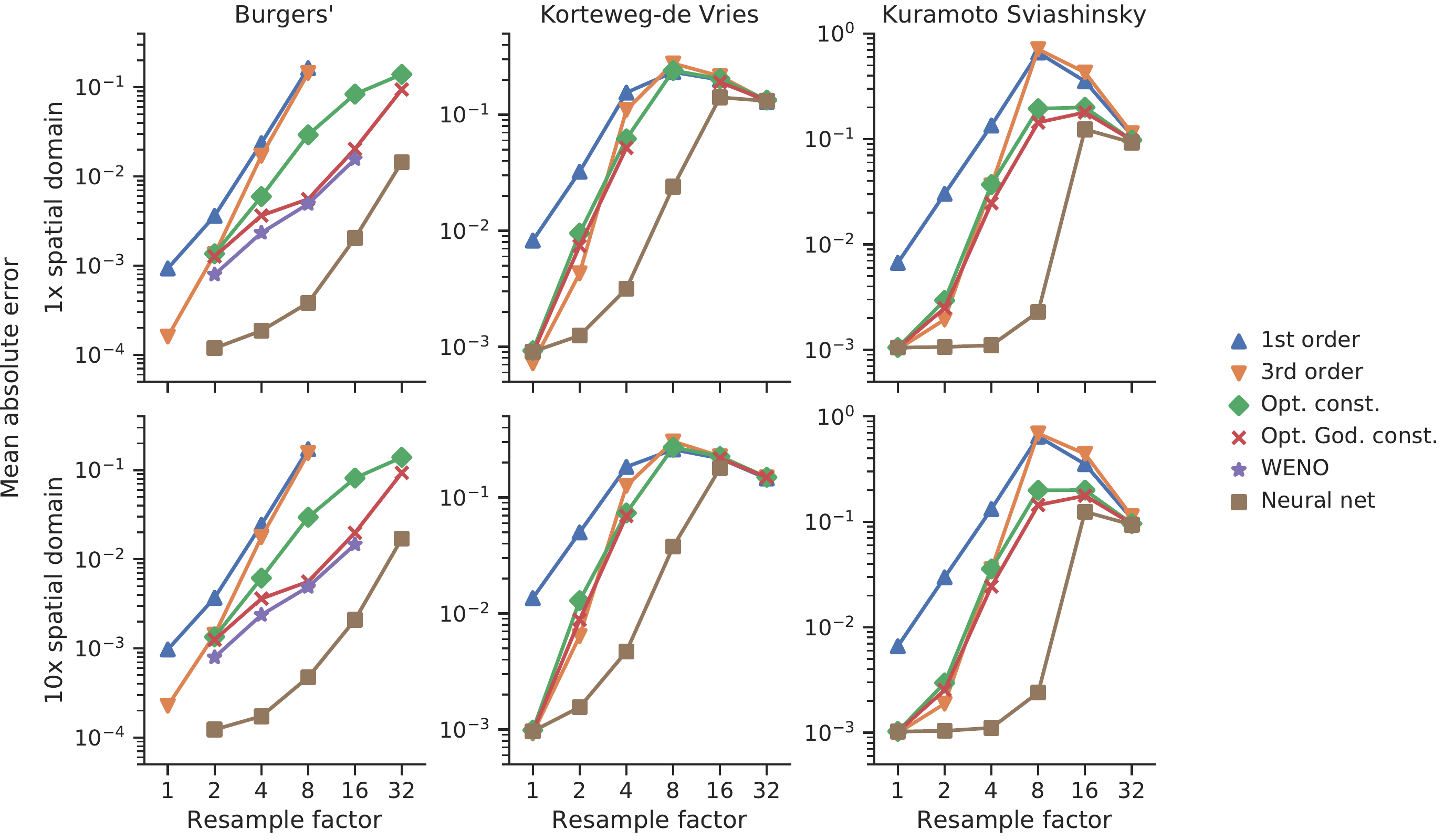}
    \caption{Mean absolute error for Burgers', KdV and KS at short times ($t \leq 15$, $t \leq 1$ and $t \leq 3$, respectively) on both the 1x spatial domain used for training and the 10x larger domain only used for inference. The 10x domain results for Burgers' equation in the bottom left panel are duplicated for comparison purposes from Fig.~3(c) of the main text.}
    \label{fig:mae-for-all-1x-and-10x}
\end{figure*}

\begin{figure*}[tbh]
    \centering
    \includegraphics[width=6.5in]{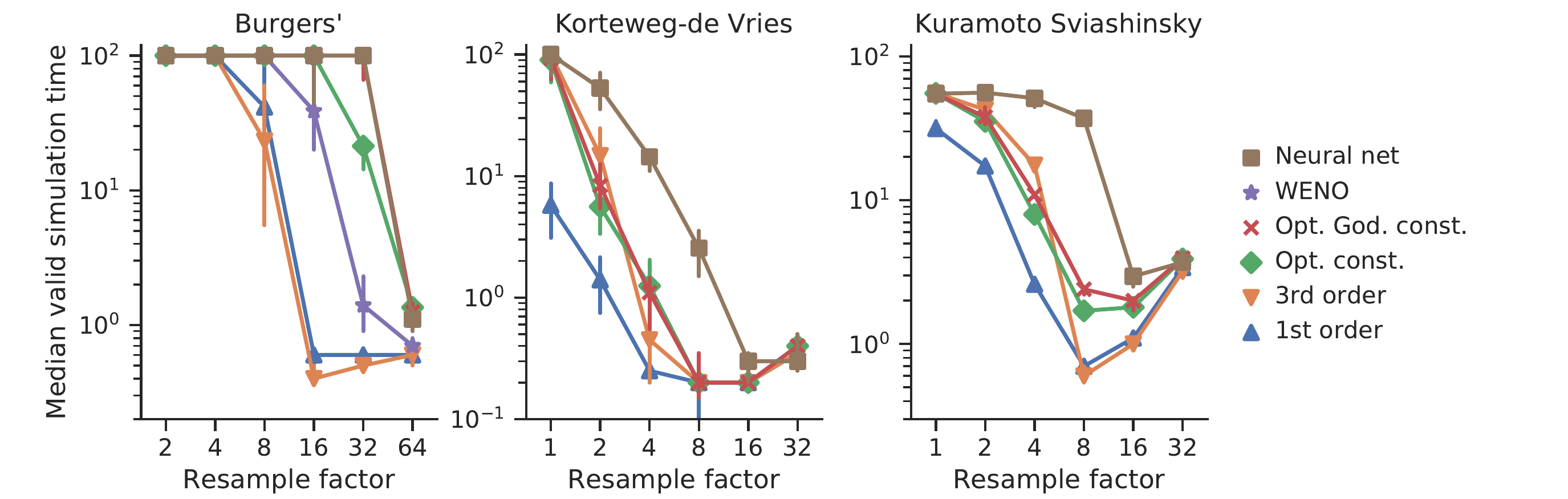}
    \caption{Survival times for Burgers', KdV and KS at short times on the 10x larger inference domain. Results for the optimized Godunov constant model for Burgers' equation are obscured in the plot by the results for the neural network model.}
    \label{fig:survival-for-all-10x}
\end{figure*}

Similar to Fig.~\ref{fig:mainresults}(c) in the main text, Fig.~\ref{fig:mae-for-all-1x-and-10x} shows mean absolute error for coarse grained models of KdV and KS at different resample factors, averaged over 100 realizations of each model on the training spatial domain, and 10 realizations of each model on a ten-times larger spatial domain.
It is only possible to compare mean absolute error for short times, because at long times bad models diverge and the mean becomes undefined.
Similar to Fig.~\ref{fig:survival_plot} in the main text, Figure~\ref{fig:survival-for-all-10x} shows median survival time for all models on the ten-times larger domain.
The ranking of models in both figures is broadly consistent with Fig.~\ref{fig:survival_plot} .

\section*{Appendix VII: Ablation study}

To understand the importance of various physical constraints in our model architecture, we performed an ablation study, comparing various modeling choices with the same neural network models. In order of increasing level of physical constraints, these include:
\begin{itemize}
\item ``Time derivative'' models that predict the time derivative $\partial u / \partial t$ directly, without any incorporation of physical prior knowledge at all.
\item ``Flux'' models that predict the flux $J$ at the boundary between grid cells and use that to compute the time derivative with the continuity equation, without any knowledge of the physical equation.
\item ``Space derivatives`` models that predict spatial derivatives without any constraints, and plug those spatial derivatives into the physical equation to compute the flux.
\item ``Coefficients'' models that predict linear coefficients used in finite difference formula for computing space derivatives.
\item ``Constrained coefficients'' models that predict linear coefficients constrained to at least first-order accuracy.
\item ``Godunov constrained coefficients'' models that predict constrained coefficients, and additionally use the Godunov numerical flux for the convective term in the flux.
\end{itemize}
In addition, we also experimented with finite-difference modeling (coarse-graining by subsampling)  rather than the finite-volume modeling (coarse-graining by averaging) used in all of the above. Generally speaking, finite difference models can also work but they perform worse than finite volumes, as with traditional numerical methods. We do not report these results here for brevity.
\begin{figure*}
    \centering
    \includegraphics[width=6.5in]{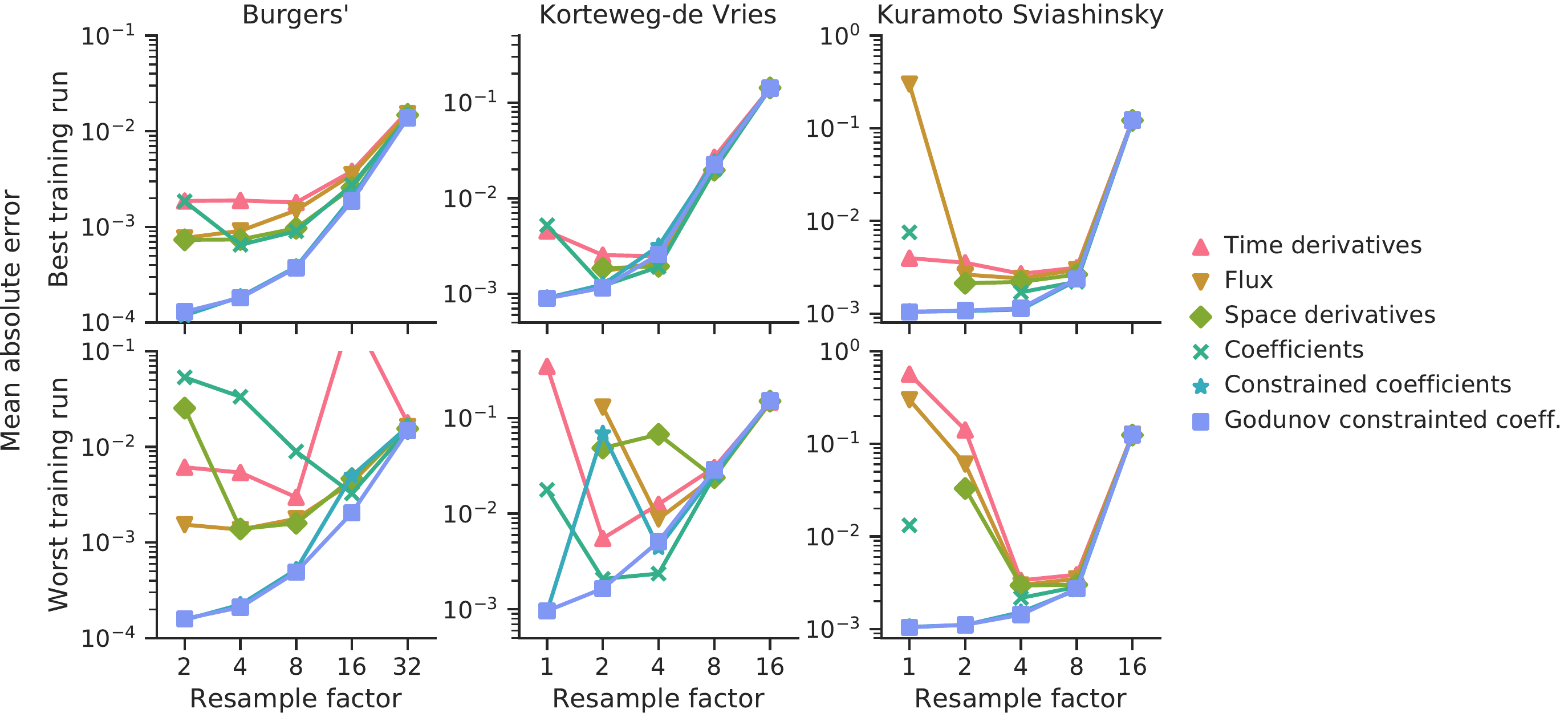}
    \vspace{0.5cm}
    \rule{6.5in}{0.5pt}
    \vspace{0.5cm}
    \includegraphics[width=6.5in]{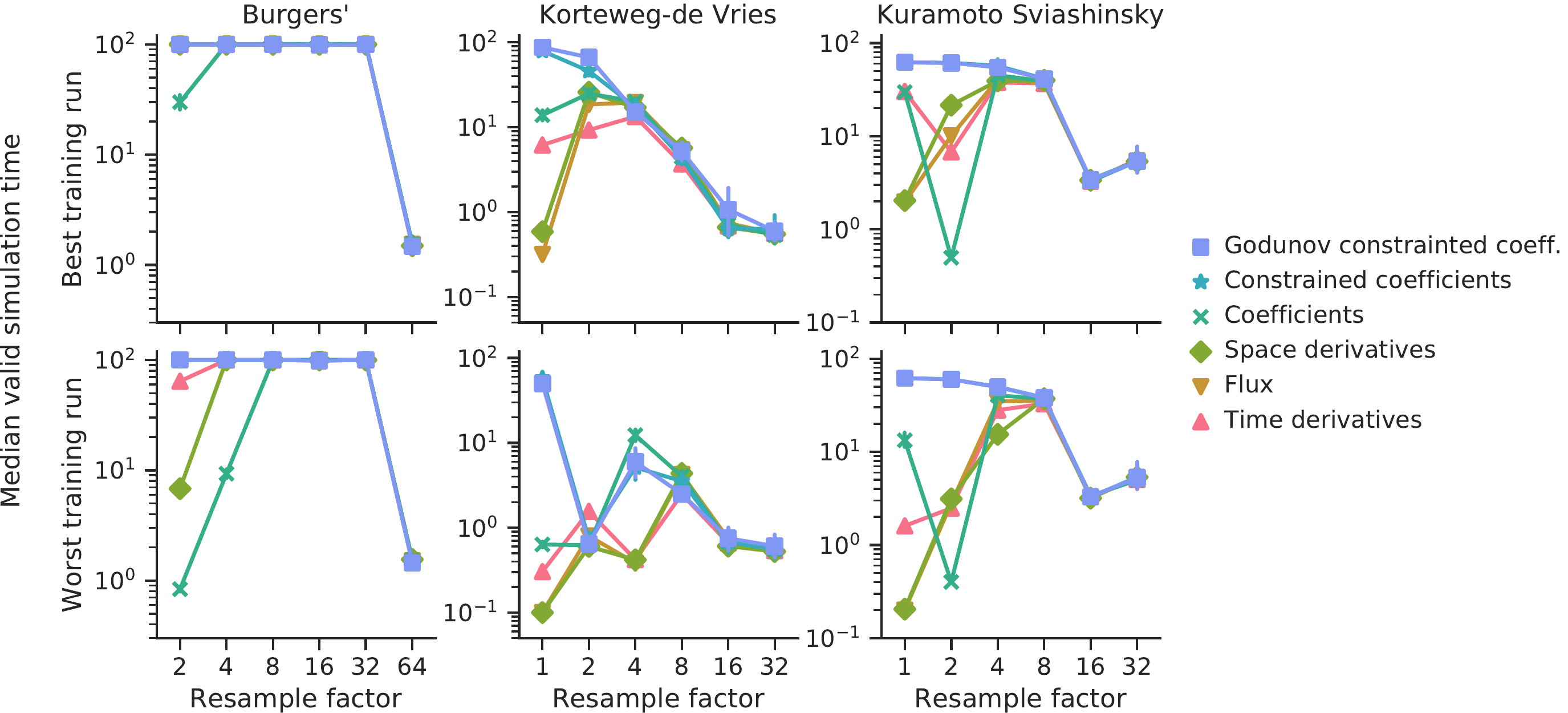}
    \caption{Mean absolute error and survival times for alternative neural network models. The ``best'' and ``worst'' training runs illustrate variability over ten random choices of initial conditions and shuffles of the training data.}
    \label{fig:ablation}
\end{figure*}

The results, shown in Fig.~\ref{fig:ablation}, suggest a number of trends:
\begin{enumerate}
\item Consistent results across all equations and grid resolutions were only achieved with the ``constrained coefficients'' and ``Godunov constrained coefficients'' models. For Burgers' and KS, these models achieved good results on all training runs.
\item Physical constraints dramatically improve the performance of trained models on fine-resolution grids, but have a smaller influence on results for low-resolution (large resample factor) grids.
This makes sense for two reasons: First, low resample factor is closer to the continuum limit, for which these physical constraints and the equations themselves hold exactly. Second,  due to the CFL condition fine-resolution models use a smaller timestep. Therefore, when integrated for the same fixed amount of time, the fine-resolution models are repeatedly applied more times than low resolution ones, making them more sensitive to numerical stability.
\item Building in the Godunov flux does not directly improve the predictions of neural network models. Apparently \eqref{eq:godunov} is simple enough that it is easy for a neural network to learn, as shown in Fig.~\ref{fig:burgers-coefficients} in the main text and Fig.~\ref{fig:burgers-coefficients-zeroth-derivative}.
\end{enumerate}

\end{document}